\documentclass[aps,pra,twocolumn,showpacs,superscriptaddress]{revtex4}
\usepackage{graphicx}
\usepackage{times}
\usepackage{amssymb}
\newcommand{\expec}[1]{\langle#1\rangle}  

\begin{document}


\title{Stokes operator squeezed continuous variable polarization states}

\author{Roman Schnabel}
\affiliation{Department of Physics, Faculty of Science, Australian
National University, ACT 0200, Australia}

\author{Warwick P. Bowen}
\affiliation{Department of Physics, Faculty of Science, Australian
National University, ACT 0200, Australia}

\author{Nicolas Treps}
\affiliation{Department of Physics, Faculty of Science, Australian
National University, ACT 0200, Australia}

\author{Timothy~C.~Ralph}
\affiliation{Department of Physics, Centre for Lasers, 
University of Queensland, St Lucia, QLD 4072, Australia}

\author{Hans-A. Bachor}
\affiliation{Department of Physics, Faculty of Science, Australian
National University, ACT 0200, Australia}

\author{Ping Koy Lam}
\affiliation{Department of Physics, Faculty of Science, Australian
National University, ACT 0200, Australia}




\date{\today}

\begin{abstract}
We investigate quantum correlations in continuous wave polarization
squeezed laser light generated from one and two optical parametric
amplifiers, respectively. A general expression of how Stokes operator
variances decompose into two mode quadrature operator variances 
is given. Stokes parameter variance spectra for four different polarization 
squeezed states have been measured and compared with a coherent state. 
Our measurement results are visualized by three-dimensional Stokes operator 
noise volumes mapped on the quantum Poincar\'{e} sphere. 
We quantitatively compare the channel capacity of the different continuous 
variable polarization states for communication protocols.  It is shown that 
squeezed polarization states provide 33\% higher channel capacities than 
the optimum coherent beam protocol.
\end{abstract}

\pacs{42.50.Dv, 42.65.Yj, 03.65.-w, 03.67.-a}
\maketitle

\section{introduction}

The quantum properties of the polarization of continuous wave light
are of increasing interest since they offer new opportunities for
communicating quantum information with light and for transferring
quantum information from atoms to photons and vice versa.  In the
single photon regime the quantum polarization states have been 
vigorously studied, theoretically and experimentally, 
with investigations of fundamental problems of
quantum mechanics, such as Bell's inequality \cite{Bell-theo,Bell-exp}, 
and of potential applications such as quantum cryptography 
\cite{QCr-theo,QCr-exp}.  
In comparison, continuous variable quantum polarization states have
received little attention.  Recently however, due to their apparent
usefulness to quantum communication schemes, interest in them has been
growing and a number of theoretical papers have been published
\cite{APu89}--\nocite{COP93,AAC98,KCh96,CAA97,ALPA02,Ralph00}\cite{KLLRS02}.

Continuous variable quantum polarization states can be carried by a
bright laser beam, providing high bandwidth capabilities and therefore
faster signal transfer rates than single photon systems.  In addition,
several proposals have been made for quantum networks that consist of
spatially separated nodes of atoms whose spin states enable the
storage and processing of information, connected by optical quantum
communication channels
\cite{DiVin95}--\nocite{CIRAC95}\cite{KuzPolzik00}.  Mapping of
quantum states from photonic to atomic media is a crucial element in
these networks.  For continuous variable polarization states this mapping 
has been experimentally
demonstrated \cite{HSSP99}.  
Even one of the fundamental phenomena of quantum physics, entanglement, has 
been realized for the macroscopic spin states of two gas samples \cite{JKP01}. 
Very recently entanglement was experimentally
demonstrated for optical continuous variable polarization states for the first
time \cite{BTSL02}.  

Several methods for generating continuous variable polarization
squeezed states have been proposed, most using nonlinearity provided
by Kerr-like media and optical solitons
\cite{KCh96,AAC98,KLLRS02}.  The two experimental demonstrations
previous to our work reported here and in \cite{BSBL02}, however, were
achieved by combining a dim quadrature squeezed beam with a bright
coherent beam on a polarizing beam splitter \cite{HSSP99,GSYL87}.  
In both cases only
the properties of the state relevant to the experimental outcome were
characterized.  The full characterization of a continuous variable
polarization state requires measurements of the fluctuations in both
the orientation, and the length of the Stokes vector on a Poincar\'{e}
sphere. 

In this paper we present the complete experimental
characterization of the Stokes vector fluctuations for four different
quantum polarization states. 
We make use of ideas recently published by Korolkova {\it et al.} \cite{KLLRS02}.
Their concept of squeezing more than one Stokes operator of a laser beam and 
a simple scheme to measure the Stokes operator variances are realized.
Our results given in
\cite{BSBL02} are extended and discussed in more detail. Experimental 
data from polarization squeezed states generated from a single 
quadrature squeezed beam and from two quadrature squeezed beams are compared.

The outline of this paper is as follows. We present a description 
of the theory involved in our experiments. Since polarization states 
can be decomposed into two mode quadrature states a general link between 
Stokes operator variances and quadrature variances is given. In the 
experimental section we characterize the polarization
fluctuations of a single amplitude squeezed beam from an optical 
parametric amplifier (OPA). It can be seen that
only the fluctuations of the Stokes vector length are below that of a
coherent beam (ie. squeezed).  Grangier {\it et al.} \cite{GSYL87}
and Hald {\it et al.} \cite{HSSP99} converted this to squeezing of the
Stokes vector orientation by combining the quadrature squeezed beam
with a much brighter coherent beam on a polarizing beam splitter.  We
experimentally generate this situation and indeed show that the Stokes
vector orientation is squeezed. This result is compared with measurements 
on polarization states generated from two quadrature squeezed beams. 
Two bright amplitude or phase squeezed beams from 
two independent OPAs are overlapped on a polarizing beam splitter 
\cite{KLLRS02,BSBL02} demonstrating ``pancake-like'' and
``cigar-like'' uncertainty volumes on the Poincar\'{e} sphere for
phase and amplitude squeezed input beams, respectively.  Both the
orientation and the length of the Stokes vector were squeezed for 
the ``cigar-like'' uncertainty volume.
In the final section several schemes for encoding information on continuous variable polarization states of light are discussed.  The conventional fiber-optic communication protocol is compared with optimized coherent beam and squeezed beam protocols.  We show that the channel capacity of the ``cigar-like'' polarization squeezed states exceeds the channel capacity of all the other states.

\section{Theoretical background}
\label{theo}
%
%
\begin{figure}[h!]
\centerline{\includegraphics[width=8.0cm]{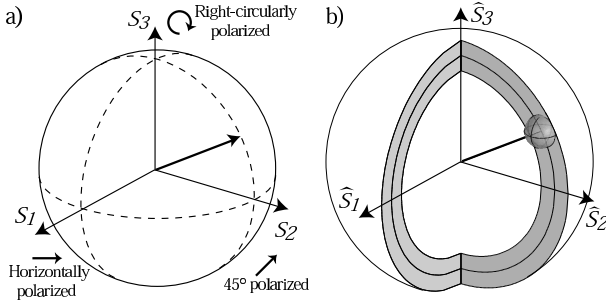}}.
\vspace{2mm}
\caption{Diagram of a) classical and b) quantum Stokes vectors
   mapped on a Poincar\'{e} sphere; the ball at the end of the quantum
   vector visualizes the quantum noise in $\rm \hat S_{1}$, $\rm \hat
   S_{2}$, and $\rm \hat S_{3}$; and the non-zero quantum sphere
   thickness visualizes the quantum noise in $\rm \hat S_{0}$.}
\label{Poincare}
\end{figure}
The polarization state of a light beam in classical optics can be
visualized as a Stokes vector on a Poincar\'{e} sphere (Fig.~1) and is
determined by the four Stokes parameters \cite{Sto52}: $S_{0}$
represents the average beam intensity whereas $S_{1}$, $S_{2}$, and $S_{3}$
characterize its polarization and form a Cartesian axes system.  If
the Stokes vector points in the direction of $S_{1}$, $S_{2}$, or
$S_{3}$ the polarized part of the beam is horizontally, linearly at
45$^\circ$, or right-circularly polarized, respectively.  Two beams
are said to be opposite in polarization and do not interfere if their
Stokes vectors point in opposite directions.  The quantity $\,S =
(S_{1}^{2}+S_{2}^{2}+S_{3}^{2})^{1/2}\,$ is the radius of the
classical Poincar\'{e} sphere and describes the average intensity of the
polarized part of the radiation.
The fraction $\,S / S_{0}\,$
($\,0\!<\!S/S_{0}\!<\!1\,$) is called the degree of polarization.  For
quasi-monochromatic laser light which is almost completely polarized
$S_{0}$ is a redundant parameter, completely determined by the other
three parameters ($S_{0}\!=\!S$ in classical optics).  All four Stokes
parameters are accessible from the simple experiments shown in
Fig.~2.\\
%
%
\begin{figure}[h!]
   \centerline{\includegraphics[width=8.0cm]{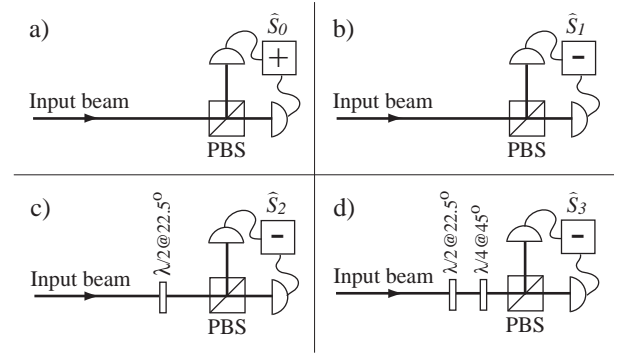}}
   \vspace{0mm}
   \caption{Apparatus required to measure each of the Stokes parameters.
   PBS: polarizing beam splitter, $\lambda/2$ and $\lambda/4$:
   half- and quarter-wave plates respectively, the plus and
   minus signs imply that an electrical sum or difference has been
   taken.}
     \label{MeasuringStokes}
\end{figure}
An equivalent representation of polarization states of light is given by
the 4 elements of the coherence matrix (Jones matrix). The relations between
these elements and the Stokes parameters can be found in \cite{WolfMandel}.
In contrast to the coherence matrix elements the Stokes parameters are
observables and therefore can be associated with Hermitian operators.
Following \cite{JauchRohrlich76} and \cite{Robson74} we define the
quantum-mechanical analogue of the classical Stokes parameters for pure
states in the commonly used notation:
\begin{eqnarray}
\label{stokes}\nonumber
\hat S_{0}\!&=& \hat a_{H}^{\dagger} \hat a^{ }_{H} + \hat a_{V}^{\dagger}
\hat a_{V}^{ } ~,\hspace{10mm}\\
\hat S_{1}\!&=& \hat a_{H}^{\dagger} \hat
a_{H}^{ } - \hat a_{V}^{\dagger} \hat a_{V}^{ } ~,\hspace{10mm}\\ \nonumber
\hat S_{2}\!&=&
\hat a_{H}^{\dagger} \hat a_{V}^{ } e^{i\theta} \!+ \hat a_{V}^{\dagger}
\hat a_{H}^{ } e^{-i\theta} ~,\\ \nonumber
\hat S_{3}\!&=& 
i\hat a_{V}^{\dagger} \hat a_{H}^{ } e^{-i\theta} \!
-i\hat a_{H}^{\dagger} \hat a_{V}^{ } e^{i\theta} ~,
\end{eqnarray}
where the subscripts $H$ and $V$ label the horizontal and vertical
polarization modes respectively, $\theta$ is the phase shift between
these modes, and the $\hat a^{ }_{H,V}$ and
$\hat a_{H,V}^{\dagger}$ are annihilation and creation operators for the
electro-magnetic field in frequency space \cite{footnote1}.

The commutation relations of the annihilation and creation operators
%
\begin{equation}
[\hat a_{k}, \hat a_{l}^{\dagger}] = \delta_{kl}
~,\hspace{4mm}\mbox{with}\hspace{4mm} k,l \in\{H,V\} ~,
\label{acomrel}
\end{equation}
directly result in Stokes operator commutation relations,
\begin{equation}
[\hat S_{1}, \hat S_{2}] =  2 i \hat S_{3} ~,\hspace{2mm}
[\hat S_{2}, \hat S_{3}] =  2 i \hat S_{1} ~,\hspace{2mm}
[\hat S_{3}, \hat S_{1}] =  2 i \hat S_{2} ~\,.
\label{Scomrel}
\end{equation}
%
Apart from a normalization factor, these relations are identical to
the commutation relations of the Pauli spin matrices.  In fact the
three Stokes parameters in Eq.  (\ref{Scomrel}) and the three Pauli
spin matrices both generate the special unitary group of symmetry
transformations SU(2) of Lie algebra \cite{Kaku93}.  Since this group
obeys the same algebra as the three-dimensional rotation group,
distances in three dimensions are invariant.  Accordingly the operator
$\hat S_{0}$ is also invariant and commutes with the other three
Stokes operators ($\,[\hat S_{0}, \hat S_{j}] = 0,
\;\mbox{with}\;j=1,2,3$).  
The non-commutability of the Stokes operators $\hat S_{1}$, $\hat S_{2}$ and
$\hat S_{3}$ precludes the simultaneous exact measurement of their physical
quantities. As a direct consequence of Eq.(\ref{Scomrel}) the Stokes operator
mean values $\,\expec{\hat S_{j}}\,$ and their variances
$\,V_j=\expec{\hat S_{j}^2}\!-\!\expec{\hat S_{j}}^2\,$ are restricted by the
uncertainty relations~\cite{JauchRohrlich76}
\begin{equation}
V_1 V_2 \ge |\expec{\hat S_{3}}|^2 ~, ~ V_2 V_3 \ge |\expec{\hat
S_{1}}|^2 ~, ~ V_3 V_1 \ge |\expec{\hat S_{2}}|^2 ~.
\label{uncer}
\end{equation}
In general this results in non-zero variances in the individual Stokes
parameters as well as in the radius of the Poincar\'{e} sphere (see
Fig.~\ref{Poincare}b)).  
The quantum noise in the Stokes parameters even effects the 
definitions of the degree of polarization \cite{APu89,AAC98} and the Poincar\'{e} 
sphere radius. It can be shown 
from Eqs.~(\ref{stokes}) and (\ref{acomrel}) that the quantum Poincar\'{e} sphere radius 
is different from its classical analogue, $\,\expec{\hat S} = \expec{\hat
S_{0}^{2} + 2 \hat S_{0}}^{1/2}\,$.

Recently it has been shown that the Stokes operator 
variances may be obtained from the frequency spectrum of the
electrical output currents of the setups shown in
Fig.~\ref{MeasuringStokes}~\cite{KLLRS02}. 
To calculate the Stokes operator variances we use the linearized 
formalism here. The creation and annihilation
operators are expressed as sums of real classical amplitudes
$\alpha^{ }_{H,V}$ and quantum noise operators $\delta \hat
a_{H,V}$ \cite{WallsMilburn}
\vspace{-1mm}
\begin{equation}
\hat a^{ }_{H,V} = \alpha^{ }_{H,V} + \delta\hat a_{H,V} ~.
\label{lin}
\end{equation}
The operators in Eq.(\ref{lin}) are non-hermitian and therefore 
non-physical. To express the Stokes operators of Eq.~(\ref{stokes}) in
terms of hermitian operators we define the generalized quadrature 
quantum noise operators $\delta \hat X_{H,V}(\xi)$
\vspace{-1mm}
\begin{eqnarray}
\delta\hat X^{ }_{H,V}(\xi)=& \delta\hat a^{\dagger}_{H,V}e^{ i\xi} 
                         + \delta\hat a^{       }_{H,V}e^{-i\xi} ~,\hspace{6mm}\\
\delta\hat X^{ }_{H,V}(\xi\!=\!0) =& 
\delta\hat X^{+}_{H,V} =\delta\hat a^{\dagger}_{H,V} + \delta\hat a^{ }_{H,V} ~,\hspace{3mm}\\
\delta\hat X^{ }_{H,V}(\xi\!=\!\pi/2) =& 
\delta\hat X^{-}_{H,V} =i(\delta\hat a^{\dagger}_{H,V} - \delta\hat a^{ }_{H,V}) ~.
\label{quad}
\end{eqnarray}
$\xi$ is the phase of the quantum mechanical oscillator and
$\delta \hat X_{H,V}^{+}$ and $\delta \hat X_{H,V}^{-}$ are the 
amplitude quadrature noise operator and the phase quadrature noise 
operator respectively.

If the variances of the noise operators are much smaller than
the coherent amplitudes then a first order approximation of the noise
operators is appropriate. This yields the Stokes operator mean values
\begin{eqnarray}
\label{expec}\nonumber
\expec{\hat S_{0}}
&=& \alpha_{H}^2+\alpha_{V}^2 = \expec{\hat n}~,\hspace{2mm}\\ 
\expec{\hat S_{1}}
&=&\alpha_{H}^2-\alpha_{V}^2 ~,\hspace{11.5mm}\\ \nonumber
\expec{\hat S_{2}}
&=& 2 \alpha^{ }_{H} \alpha^{ }_{V} \,\mbox{cos}\theta~,\\ \nonumber
\expec{\hat S_{3}}
&=& 2 \alpha^{ }_{H} \alpha^{ }_{V} \,\mbox{sin}\theta ~,
\end{eqnarray}
These expressions are identical to
the Stokes parameters in classical optics.  Here $\,\expec{\hat n}\,$
is the expectation value of the photon number operator.  For a
coherent beam the expectation value and variance of $\hat n$ have the
same magnitude, this magnitude equals the conventional shot-noise
level.  The variances of the Stokes parameters are given by
\begin{eqnarray}
\,V_0=&\!\alpha_{H}^2 \expec{(\delta\hat X_{H}^+)^2}\!
    + \alpha_{V}^2 \expec{(\delta\hat X_{V}^+)^2}
    +2\alpha_{H}^{}\alpha_{V}^{}\expec{\delta\hat X_{H}^+\delta\hat X_{V}^+}
,\nonumber\hspace{5mm}\\[1.9mm]
\,V_1=&\!\alpha_{H}^2 \expec{(\delta\hat X_{H}^+)^2}\!
    + \alpha_{V}^2 \expec{(\delta\hat X_{V}^+)^2}
    -2\alpha_{H}^{}\alpha_{V}^{}\expec{\delta\hat X_{H}^+\delta\hat X_{V}^+}
,\nonumber\hspace{5mm}
\end{eqnarray}

\vspace{-7mm}
\begin{eqnarray}
\label{var}
V_2(\theta)&\!\!=\alpha_{H}^2 \expec{(\delta\hat X_{V}^{}(-\theta))^2}\!
    + \alpha_{V}^2 \expec{(\delta\hat X_{H}^{}(\theta))^2}
    \;,\hspace{13mm} \\
    &+2\alpha_{H}^{}\alpha_{V}^{}\expec{\delta\hat X_{V}^{}(-\theta)\delta\hat
    X_{H}^{}(\theta)}
\;,\nonumber\hspace{16mm}\\[1mm]
V_3(\theta)&\!= V_2(\theta\!-\!\frac{\pi}{2})\;.\hspace{54mm}\nonumber
%
\end{eqnarray}
It can be seen from Eqs.(\ref{var}) that the variances of
Stokes operators can be expressed in terms of the variances of quadrature
operators of two modes. The polarization squeezed state can then be
defined in a straight forward manner.
The variances of the noise operators in the above equation are
normalized to one for a coherent beam.  Therefore the variances of the
Stokes parameters of a coherent beam are all equal to the shot-noise
of the beam.  For this reason a Stokes parameter is said to be
squeezed if its variance falls below the shot-noise of an equal power
coherent beam.
Although the decomposition to the $H,V$-polarization
axis of Eqs.~(\ref{var}) is independent of the actual procedure of
generating a polarization squeezed beam, it becomes clear that two
overlapped quadrature squeezed beams can produce a single polarization
squeezed beam.
If two beams in the horizontal and vertical polarization mode
having uncorrelated quantum noise are used Eqs.~(\ref{var}) can be rewritten as
\begin{eqnarray}
\label{var2}
V_0\,=\!\!&\!\!\!\! V_1 \,= \alpha_{H}^2 \expec{(\delta\hat X_{H}^+)^2}\! +
\!\alpha_{V}^2 \expec{(\delta\hat X_{V}^+)^2}
~,\hspace{1mm}\nonumber\\[1mm]
V_2(\theta)=\!\!&\!\!\!
   \mbox{cos}^2\theta
\left(\alpha_{V}^2 \expec{(\delta\hat X_{H}^+)^2}\! + \!\alpha_{H}^2
\expec{(\delta \hat X_{V}^+)^2}\right) \hspace{3mm}\nonumber\\
   \!&+\mbox{sin}^2\theta
\left(\alpha_{V}^2 \expec{(\delta\hat X_{H}^-)^2}\! + \!\alpha_{H}^2
\expec{(\delta \hat X_{V}^-)^2}\right) ,\hspace{5mm}\\[1mm]
V_3(\theta) =\!\!&\! V_2(\theta\!-\!\frac{\pi}{2}) ~.\hspace{42mm}
\nonumber
\end{eqnarray}
Here we choose the amplitude and the phase quadrature noise
operators to express the variances. This corresponds to our actual
experimental setup where either amplitude or the phase quadratures
were squeezed.
It can be seen from Eqs.(\ref{var2}) that in a polarization squeezed beam
generated from two amplitude squeezed beams $\hat S_0$ and two 
additional Stokes
parameters can in theory be perfectly squeezed while the fourth is
anti-squeezed if specific angles of $\,\theta \!=\!
0$ or $\,\theta \!=\!\pi/2\,$ are used. Utilizing
only one squeezed beam it is not possible to simultaneously squeeze
any two of $\hat S_{1}$, $\hat S_{2}$, and $\hat S_{3}$ to quieter
than 3~dB below shot-noise ($V_{i} \!  + \!  V_{j} \!  \ge \!
\expec{\hat n}$, with $i,j \in \{1,2,3; i \!  \not= \!  j\}$).

\section{Experiment}\label{Experiment}
\begin{figure}[h!]
  \centerline{\includegraphics[width=8.0cm]{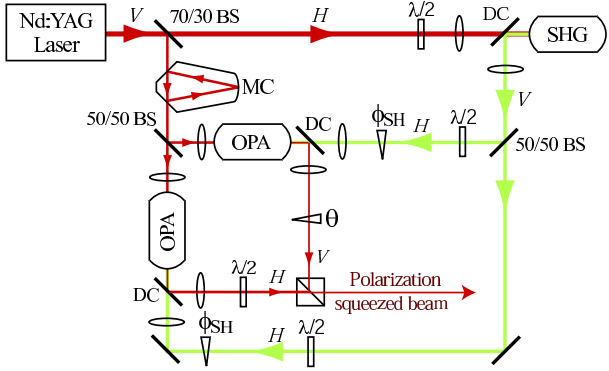}}
  \vspace{0mm}
  \caption{Schematic of the polarization squeezing experiment.  MC:
	mode cleaner, BS: beam
	splitter, DC: dichroic beam splitter, $\lambda/2$: half-wave plate,
	$\phi_{\rm{SH}}$: phase shift between 532~nm and 1064~nm light at the OPAs,
	$\theta$: phase shift between quadrature squeezed beams, PBS:
	polarizing beam splitter, H: horizontal polarization mode, V: vertical
	polarization mode.}
  \label{ExpPolSQZ}
\end{figure}

Prior to our work presented here and in \cite{BSBL02}, polarization
squeezed states were generated by combining a strong coherent beam
with a single weak amplitude squeezed beam \cite{GSYL87,HSSP99}.  In
both of those experiments the variance of only one Stokes parameter
was determined, and therefore the polarization state was not fully
characterized.  In this paper we experimentally characterize the mean
and variance of all four Stokes operators for these states.  We extend
the work to polarization squeezed states produced from two
amplitude/phase squeezed beams.  Fig.~\ref{ExpPolSQZ} shows our
experimental setup.

\subsection{Generation of quadrature squeezed light}\label{sqz}
\begin{figure}[h!]
  \centerline{\includegraphics[width=8.0cm]{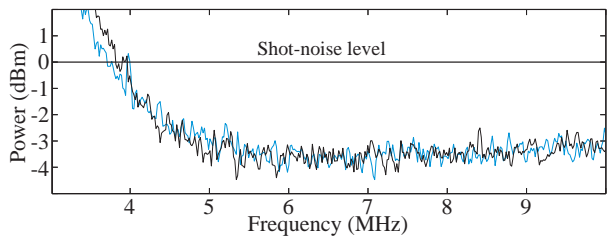}}
  \vspace{0mm}
  \caption{Typical measured variance spectra of the two locked bright quadrature squeezed beams.}
  \label{individualSQZ}
\end{figure}

We produced the two quadrature squeezed beams required for this
experiment in a pair of OPAs.  Each OPA was an optical resonator
consisting of a hemilithic MgO:LiNbO$_{3}$ crystal and an output
coupler.  The reflectivities of the output coupler were 96\% and 6\%
for the fundamental (1064~nm) and the second harmonic (532~nm) laser
modes, respectively.  Each OPA was pumped with single-mode 532~nm
light generated by a 1.5~W Nd:YAG non-planar ring laser and frequency
doubled in a second harmonic generator (SHG).  The SHG was of
identical structure to the OPAs but with 92\% reflectivity at 1064~nm.
The OPAs were seeded with 1064~nm light after spectral filtering in a
modecleaner.  The refractive indices of the MgO:LiNbO$_{3}$ crystals
in each resonator was modulated with an RF field, this provided error
signals on the reflected seed power that were used to
lock their lengths.  The modulation also resulted in a phase modulation
on the output beams from the SHG and each OPA. The coherent amplitude
of each OPAs output was a deamplified/amplified version of the seed
coherent amplitude; the level of amplification was dependent on the
phase difference between pump and seed ($\phi_{\rm{SH}}$).  Therefore the
second harmonic pump phase modulation resulted in a modulation of the
amplification of the OPAs. Error signals could be extracted from this
effect, enabling the relative phase between pump and seed to be
locked.  Locking to deamplification or amplification provided an
amplitude or phase squeezed output, respectively. 
Typical measured 
variance spectra of the two locked quadrature squeezed beams are shown 
in Fig.~\ref{individualSQZ}. Since the squeezed states were carried by 
bright laser beams of approximately 1~mW, the noise reduction was degraded at 
lower frequencies due to the laser relaxation oscillation. At higher 
frequencies the squeezed spectrum was limited by the bandwidth of the OPAs.
%

Past experiments requiring two quadrature squeezed beams commonly used
a single ring resonator with two outputs \cite{Ou}; with two
independent OPAs the necessary intracavity pump power is halved, this
reduces the degradation of squeezing due to green-induced infrared
absorption (GRIIRA) \cite{Furukawa}.

\subsection{Measuring the Stokes operators}\label{measuringStokes}
Instantaneous values of the Stokes operators of all polarization
states analyzed in this paper were obtained with the apparatus
shown in Fig.~\ref{MeasuringStokes}. The uncertainty
relations of Eqs.~\ref{uncer} dictate that $\hat S_{1}$, $\hat S_{2}$,
and $\hat S_{3}$ cannot be measured simultaneously.  The beam under
interrogation was split on a polarizing beam splitter and the two
outputs were detected on a pair of high quantum efficiency photodiodes
with 30~MHz bandwidth; the resulting photocurrents were added and
subtracted to yield photocurrents containing the instantaneous values
of $\hat S_{0}$ and $\hat S_{1}$.  To measure $\hat S_{2}$ the
polarization of the beam was rotated by 45$^\circ$ with a half-wave
plate before the polarizing beam splitter and the detected
photocurrents were subtracted.  To measure $\hat S_{3}$ the
polarization of the beam was again rotated by 45$^\circ$ with a
half-wave plate and a quarter-wave plate was introduced before the
polarizing beam splitter such that a horizontally polarized input beam
became right-circularly polarized.  Again the detected photocurrents were
subtracted.  The expectation value of each Stokes operator was equal
to the DC output of the detection device and the variance was obtained
by passing the output photocurrent into a Hewlett-Packard E4405B
spectrum analyzer.  Every polarization state interrogated in this work
had a total power $\expec{\hat S_{0}}$ of roughly 2~mW.

An accurate shot-noise level was required to determine whether any
given Stokes operator was squeezed.  This was measured by operating a
single OPA without the second harmonic pump.  The seed power was
adjusted so that the output power was equal to that of the beam being
interrogated.  In this configuration the detection setup for $\hat
S_{2}$ (see Fig.~\ref{MeasuringStokes}c)) functions exactly as a
homodyne detector measuring vacuum noise scaled by the OPA output
power, the variance of which is the shot-noise.  Throughout each
experimental run the power was monitored and was always within 2\% of
the power of the coherent calibration beam.  This led to a
conservative error in our frequency spectra of $\pm$0.05~dB.

The Stokes operator variances reported in this paper were taken over
the range from 3 to 10~MHz.  The darknoise of the detection apparatus
was always more than 4~dB below the measured traces and was taken into
account.  Each displayed trace is the average of three measurement
results normalized to the shot-noise and smoothed over the resolution
bandwidth of the spectrum analyzer which was set to 300~kHz.  The
video bandwidth of the spectrum analyzer was set to 300~Hz.

As is the case for all continuous variable quantum optical
experiments, the efficiency of the Stokes operator measurements was
critical.  The overall detection efficiency of the interrogated beams
was 76\%.  The loss came primarily from three sources: loss in escape
from the OPAs (14\%), detector inefficiency (7\%), and loss in optics
(5\%).  In the experiment where a squeezed beam was overlapped with a
coherent beam additional loss was incurred due to poor mode-matching
between the beams and the detection efficiency was 71\%.  Depolarizing
effects are thought to be another significant source of loss for some
polarization squeezing proposals \cite{APu89}.  In our scheme the
non-linear processes (OPAs) are divorced from the polarization
manipulation (wave plates and polarizing beam splitters), and
depolarizing effects are insignificant.

\subsection{Quantum polarization states from a single squeezed
beam}\label{exp1}
\begin{figure}[th]
  \centerline{\includegraphics[width=7cm]{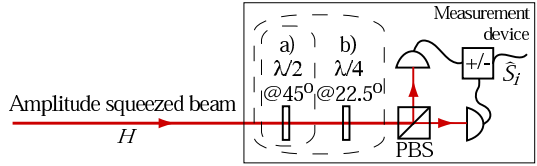}}
  \vspace{8mm}
  \caption{Apparatus used to produce and analyze a single amplitude
	squeezed beam.  Optics in a) and b) were included to measure 
      the variance and the expectation value of $\hat S_{2}$
	and $\hat S_{3}$, respectively.}
  \label{OneSQZExpt}
\end{figure}
\begin{figure}[h]
  \centerline{\includegraphics[width=8.0cm]{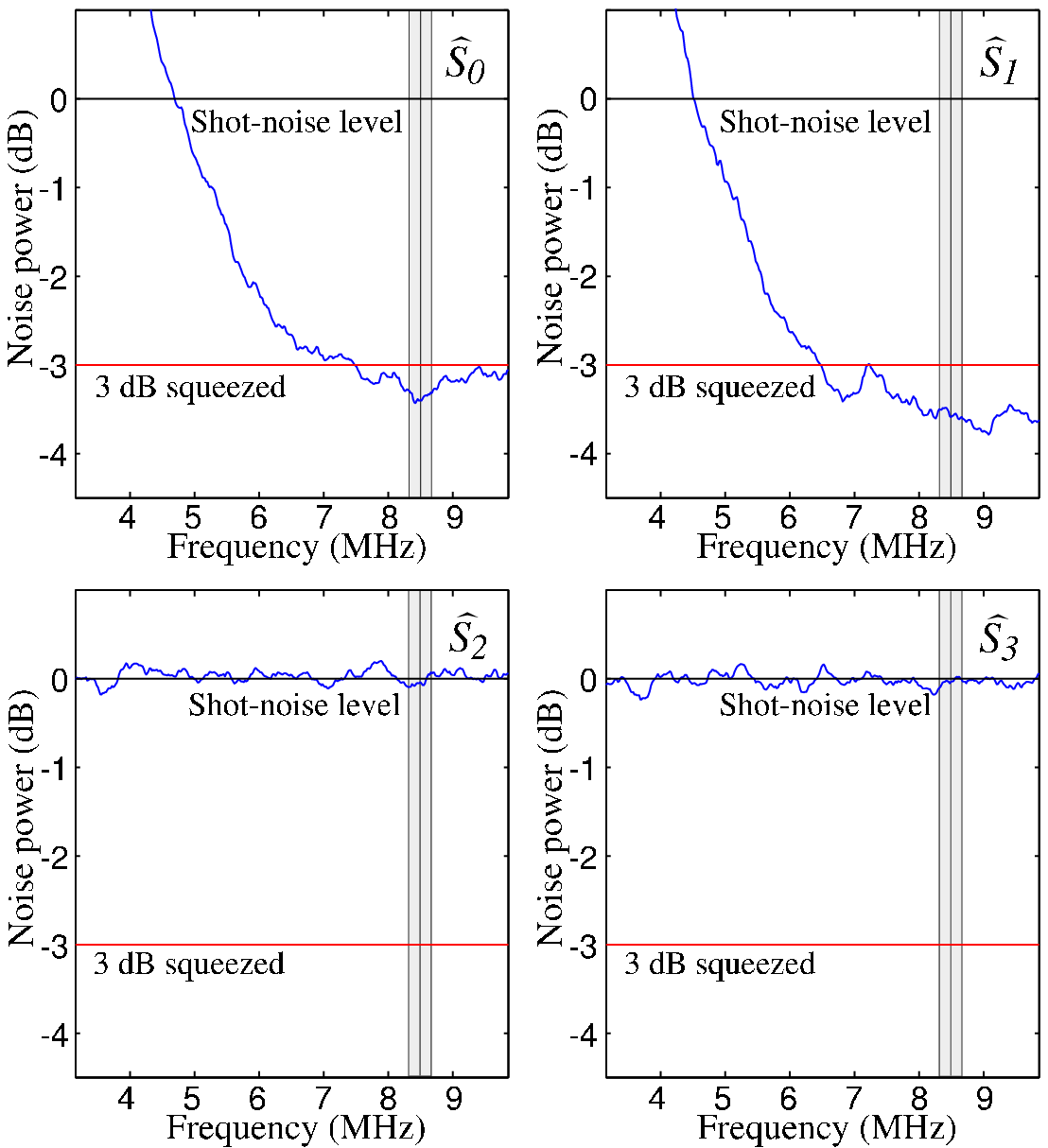}}
  \vspace{1mm}
  \caption{Measured variance spectra of quantum noise on $\hat
  	S_{0}$, $\hat S_{1}$, $\hat S_{2}$, and $\hat S_{3}$ for
  	a single bright amplitude squeezed beam; normalized to shot-noise.
  	The shaded region was used to construct the Poincar\'{e} sphere
  	representation in Fig.~\ref{BallsAll} b).}
  \label{1SQZ}
\end{figure}
We first characterize the polarization state of a single bright
amplitude squeezed beam provided by one of our OPAs, as shown in
Fig.~\ref{OneSQZExpt}.  The squeezed beam was horizontally polarized,
resulting in Stokes operator expectation values of $\langle \hat S_{0}
\rangle \!=\!  \langle \hat S_{1} \rangle \!=\!  |\alpha_{H}|^{2}$ and
$\langle \hat S_{2} \rangle \!=\!  \langle \hat S_{3} \rangle \!=\!
0$.  The variance spectra of the operators were measured and are
displayed in Fig.~\ref{1SQZ}.  $\hat S_{0}$ and $\hat S_{1}$ were
squeezed since the horizontally polarized amplitude squeezed beam hit only one
detector in this detector setup.  
For the measurements of $\hat S_{2}$ and $\hat S_{3}$ the
beam intensity was divided equally between the two detectors.  The
electronic subtraction yielded vacuum noise scaled by the beam
intensity, thus both variance measurements were at the shot noise
level.  It is apparent from these measurements that only the length of
the Stokes vector is well determined; the orientation is just as
uncertain as it would be for a coherent state.
\begin{figure}[h]
  \centerline{\includegraphics[width=8.3cm]{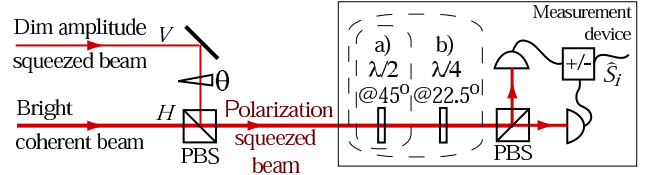}}
  \vspace{0mm}
  \caption{Apparatus used to produce and analyze the polarization
	squeezed beam produced by overlapping a dim quadrature squeezed beam
	with a bright coherent beam.  Optics in a) and b) were included to measure 
      the variance and the expectation value of $\hat S_{2}$
	and $\hat S_{3}$, respectively.}
  \label{OneSQZonecohExpt}
\end{figure}
\begin{figure}[h]
  \centerline{\includegraphics[width=8.0cm]{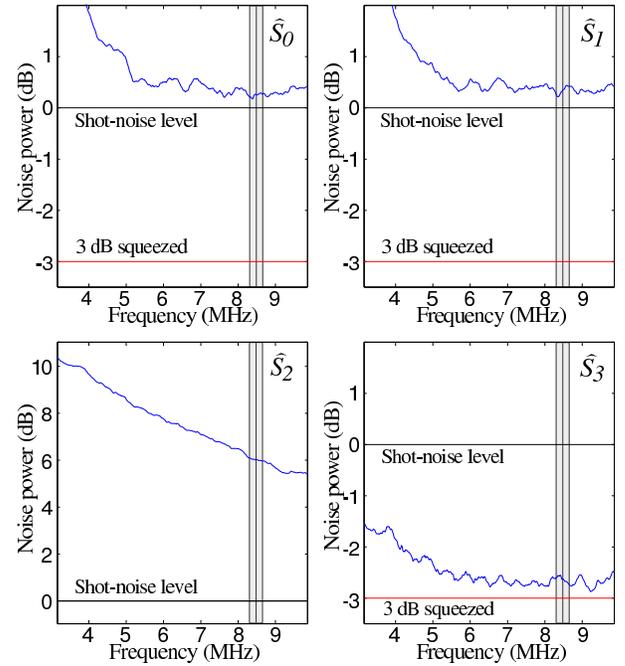}}
  \vspace{0mm}
  \caption{Measured variance spectra of quantum noise on $\hat
  S_{0}$, $\hat S_{1}$, $\hat S_{2}$, and $\hat S_{3}$ for a
  vacuum amplitude squeezed and a bright coherent input beams;
  normalized to shot-noise.  The shaded region was used to construct
  the Poincar\'{e} sphere representation in Fig.~\ref{BallsAll} c).}
  \label{1SQZ1Coh}
\end{figure}
To obtain squeezing of the orientation of the Stokes vector Grangier
{\it et al.} \cite{GSYL87} and Hald {\it et al.} \cite{HSSP99}
overlapped a dim quadrature squeezed beam with a bright orthogonally
polarized coherent beam.  We consider this situation next (as shown in
Fig.  \ref{1SQZ1Coh}).  Since two beams are now involved, the relative
phase $\theta$ becomes important.  A DC and an RF error-signal, both
dependant on $\theta$, were extracted from the Stokes operator
measurement device.  Together, these error signals allowed us to lock
$\theta$ to either 0 or $\pi/2$~rads in all of the following
experiments.
We mixed a bright horizontally polarized coherent beam with a dim
vertically polarized amplitude squeezed beam.  Since the horizontally
polarized beam was much more intense than the vertically polarized
beam, the Stokes operator expectation values became $\langle \hat
S_{0} \rangle \!  \approx \!  \langle \hat S_{1} \rangle \!  \approx
\!  |\alpha_{H}|^{2}$ and $\langle \hat S_{2} \rangle \!  \approx \!
\langle \hat S_{3} \rangle \!=\!  0$.  The Stokes operator variances
obtained for this polarization state are shown in Fig.~\ref{1SQZ1Coh}, here $\hat
S_{2}$ is anti-squeezed and $\hat S_{3}$ is squeezed.  The variances
of $\hat S_{0}$ and $\hat S_{1}$ were slightly above the shot-noise
level because of residual noise from our laser resonant relaxation
oscillation.  The experiment carried out with $\theta$ locked to
0~rads is not shown, in this case the measured variances of $\hat
S_{2}$ and $\hat S_{3}$ were swapped.  In fact the Stokes vector was
still pointing along $\hat S_{1}$ but the quantum noise was rotated on
the Poincar\'{e} sphere (see Fig.~\ref{rotatefig}b).

\subsection{Quantum polarization states from two quadrature squeezed
beams}\label{exp2}
\begin{figure}[h]
  \centerline{\includegraphics[width=8.0cm]{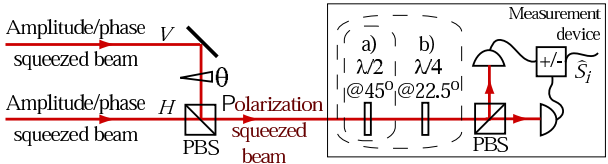}}
  \vspace{0mm}
  \caption{Apparatus used to produce and analyze the beam produced by
	combining two quadrature squeezed beams. Optics in a) and b) were 
      included to measure 
      the variance and the expectation value of $\hat S_{2}$
	and $\hat S_{3}$, respectively.}
  \label{TwoSQZExpt}
\end{figure}
\begin{figure}[h!!]
  \centerline{\includegraphics[width=8.3cm]{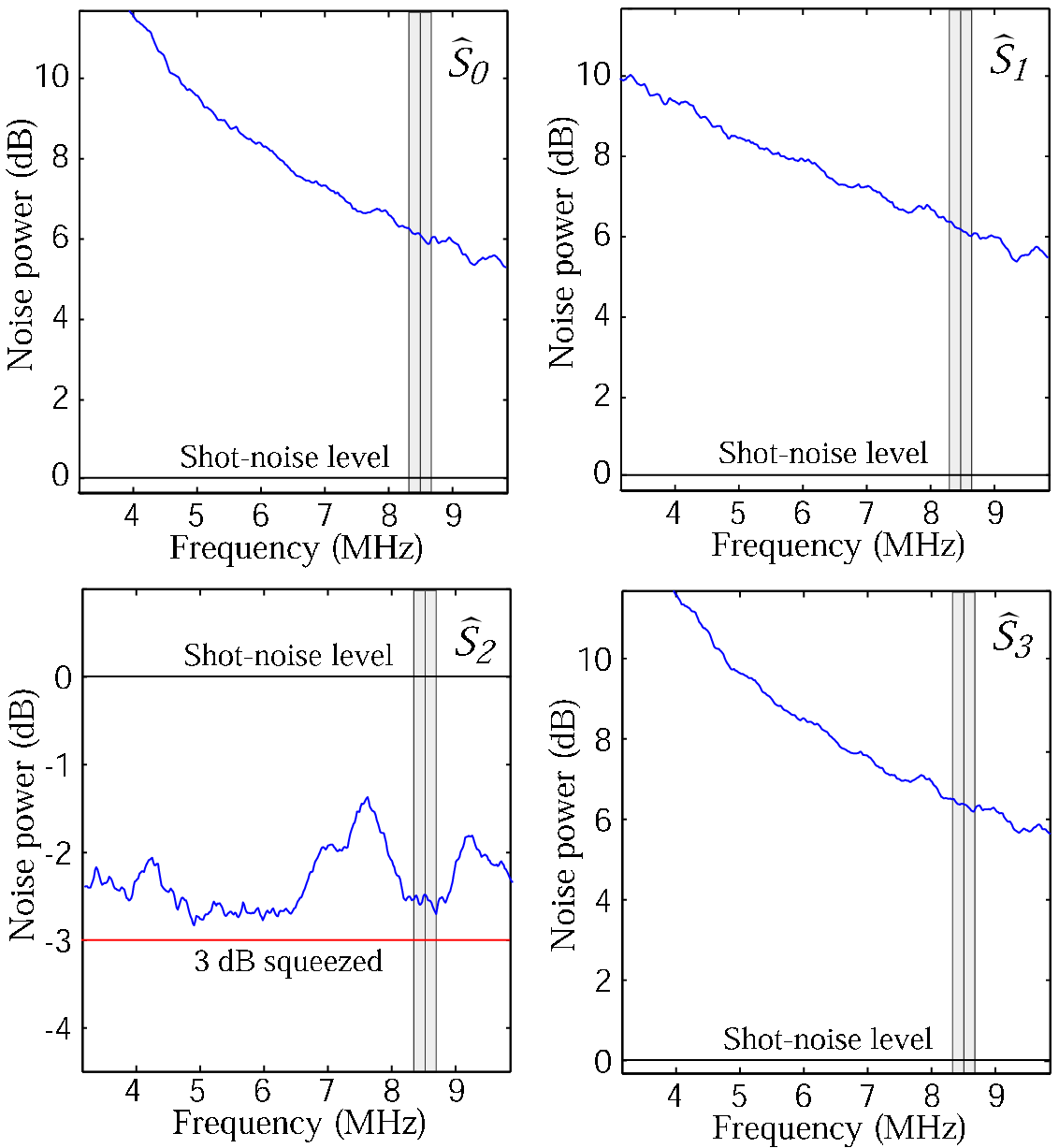}} 
  \vspace{0mm}
  \caption{Measured variance spectra of quantum noise on $\hat
  S_{0}$, $\hat S_{1}$,  $\hat S_{2}$, and $\hat S_{3}$ for two
  locked phase squeezed input beams; normalized to shot-noise.  The
  shaded region was used to construct the Poincar\'{e} sphere
  representation in Fig.~\ref{BallsAll} d).}
  \label{PhaseResults}
\end{figure}
\begin{figure}[h!!]
  \centerline{\includegraphics[width=8.0cm]{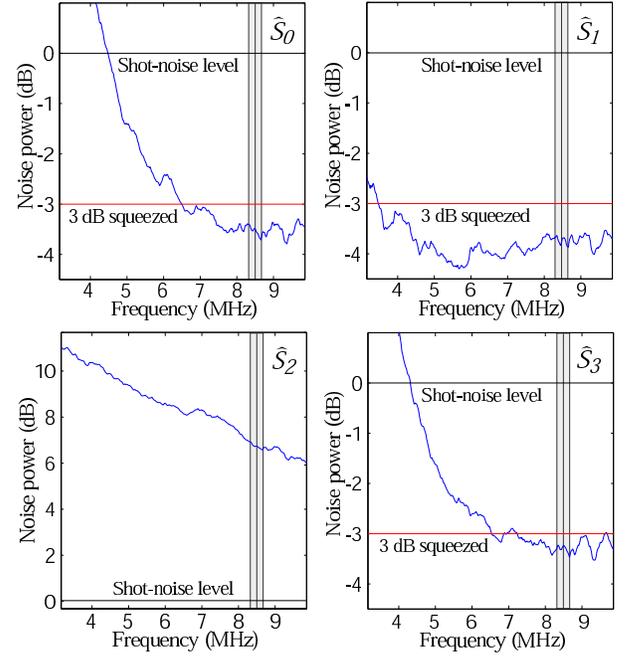}} 
  \vspace{0mm}
  \caption{Measured variance spectra of quantum noise on
  $\hat S_{0}$, $\hat S_{1}$, $\hat S_{2}$, and $\hat
  S_{3}$ for two locked amplitude squeezed input beams; normalized to
  shot-noise.  The shaded region was used to construct the Poincar\'{e}
  sphere representation in Fig.~\ref{BallsAll} e).}
  \label{AmpResults}
\end{figure}
%
%
%
%
%

%
%
%
%
The two experiments described in Section~\ref{exp1} demonstrated how it is 
possible to squeeze
the length and orientation of the Stokes vector.  In this section
we demonstrate that it is possible to do both simultaneously.  The two
quadrature squeezed beams produced in our OPAs were combined with
orthogonal polarization on a polarizing beam splitter \cite{KLLRS02}
as shown in Fig.~\ref{TwoSQZExpt}.  This produced an output beam with
Stokes parameter variances as given by Eqs.~(\ref{var2}).  Both input
beams had equal power ($\alpha_{H}^{ } \!  = \!  \alpha_{V}^{ } \!=\!
\alpha/\sqrt{2}$) and both were squeezed in the same quadrature.  The
Stokes parameters and their variances were again determined as shown
in Fig.~\ref{MeasuringStokes}.  The relative phase between the
quadrature squeezed input beams $\theta$ was locked to $\pi/2$ rads
producing a right-circularly polarized beam with Stokes parameter
means of $\langle \hat S_{1} \rangle \!=\!  \langle \hat S_{2} \rangle
\!=\!  0$ and $\langle \hat S_{0} \rangle \!=\!  \langle \hat S_{3}
\rangle \!=\!  |\alpha|^{2}$.

First both OPA pump beams were phase locked to amplification, this produced two phase
squeezed beams.  Fig.~\ref{PhaseResults} shows the measurement results
obtained; $ \hat S_{0}$, $\hat S_{1}$ and $\hat S_{3}$ were all
anti-squeezed and $\hat S_{2}$ was squeezed throughout the range of
the measurement.  The optimum noise reduction of $\hat S_{2}$ was
2.8~dB below shot-noise and was observed at 4.8~MHz.  Our OPAs are
particularly sensitive to phase noise coupling in from the
MgO:LiNbO$_{3}$ crystals.  We attribute the structure in the frequency
spectra of $\hat S_{2}$ and the poorer squeezed observed here, to
this.  Apart from this structure, these results are very
similar to those produced by a single squeezed beam and a coherent
beam; the orientation of the Stokes vector is squeezed.  However, here
the uncertainty in the length of the Stokes vector is greater than for
a coherent state so the polarization state, although produced from two
quadrature squeezed beams, is actually less certain.

Fig.~\ref{AmpResults} shows the measurement results obtained with the
OPAs locked to deamplification. Therefore both OPAs provided amplitude 
squeezed beams.
Again we interrogated the combined beams and found 
$\hat S_{0}$, $\hat S_{1}$ and $\hat S_{3}$ all to be squeezed from
4.5~MHz to the limit of our measurement, 10~MHz.  $\hat S_{2}$ was
anti-squeezed throughout the range of the measurement.  Between
7.2~MHz and 9.6~MHz $\hat S_{0}$, $\hat S_{1}$ and $\hat S_{3}$ were
all more than 3~dB below shot-noise.  The squeezing of $\hat S_{0}$
and $\hat S_{3}$ was degraded at low frequency due our lasers resonant
relaxation oscillation.  Since this noise was correlated it canceled in
the variance of $\hat S_{1}$.  The maximum squeezing of $\hat S_{0}$
and $\hat S_{2}$ was 3.8~dB and 3.5~dB respectively and was observed
at 9.3~MHz.  The maximum squeezing of $\hat S_{1}$ was 4.3~dB at
5.7~MHz.  The repetitive structure at 4, 5, 6, 7, 8 and 9~MHz was
caused by electrical pick-up in our SHG resonator emitted from a
separate experiment operating in the laboratory.  In this case both
the orientation and the length of the Stokes vector were squeezed.

Finally we point out that the variances of $\hat S_{1}$ in 
Figs.~\ref{1SQZ1Coh}, \ref{PhaseResults} and \ref{AmpResults} were 
all squeezed at frequencies down to 3~MHz and even below, whereas 
Fig.~\ref{individualSQZ} shows a clear degradation below 4~MHz. 
This improved performance is due to electrical noise cancellation of 
correlated laser relaxation oscillation noise. This noise is effectively 
reduced by taking the difference of the two photo currents in our detector 
setup to measure $V_1$.

\section{Visualization of quantum correlations in continuous
variable polarization states}
\label{VisCorrelation}
\begin{figure}[h!]
  \vspace{0mm}
  \caption{Measured quantum polarization noise at 8.5~MHz from
  different combinations of input beams.  a) single coherent beam,
  b) coherent beam and squeezed vacuum, c) bright squeezed beam, d) two
  phase squeezed beams and e) two amplitude squeezed beams.  The
  surface of the ellipsoids defines the standard deviation of the noise
  normalized to shot-noise ($\sigma_{S_{i}} \!  = \!  \sqrt{V_{i}}$).
  $\hat S_{/\!/}$ denotes the Stokes operator along the Stokes vector.}
  \label{BallsAll}
\end{figure}
\begin{figure}[h!]
  \vspace{0mm}
  \caption{Visualization of measured quantum noise and measured Stokes vectors of four polarization states mapped onto the Poincar\'{e} sphere. 
The states were generated from a) two bright amplitude squeezed inputs and b) a bright coherent beam and a amplitude squeezed vacuum. 
The rotation in a) and also in b) was achieved by a $\,\theta = \pi/2$ phase shift or an additionally introduced $\lambda/4$ wave plate.}
  \label{rotatefig}
\end{figure}

In this section measured quantum correlations in polarization states at 
8.5~MHz are visualized. 
Based on the theoretical formalism in Section~\ref{theo} 
continuous variable polarization states can be characterized by the measurement of 
Stokes operator expectation values and variances using the 
setup shown in Fig.~\ref{MeasuringStokes}. Our noise measurement results at 8.5~MHz on
five different states are summarized in Fig.~\ref{BallsAll}. 
The noise characteristics of the Stokes parameters are mapped onto the 
coordinate system of the Poincar\'{e} sphere, assuming Gaussian noise statistics.
Given this assumption, the standard deviation contour-surfaces shown
here provide an accurate representation of the states three-dimensional noise distribution.  
The quantum polarization noise of a
coherent state forms a sphere of noise as portrayed in
Fig.~\ref{BallsAll}~a). The noise volumes b) to e) visualizes the 
measurements on a single bright amplitude squeezed beam, on the combination of a 
vacuum amplitude squeezed beam and a bright coherent beam, on two locked phase 
squeezed input beams and on two locked amplitude squeezed beams, respectively. 
In all cases the 
Stokes operator noise volume describes the end position of the Stokes vector pointing upwards.
In b) and c) the Stokes vectors are parallel to the direction of $S_1$, in d) and e) 
parallel to the direction of $S_3$, since we used horizontally and right-circularly 
polarized light, respectively.
However, there was no fundamental bias in the orientation of the quantum Stokes
vector in our experiment.  By varying the angle of an additional half-wave plate in the
polarization squeezed beam or by varying $\theta$ any orientation may
be achieved.  In fact, as mentioned earlier our experiments were also
carried out with $\theta$ locked to 0 rads.  This had the effect of
rotating the Stokes vector and its quantum noise by $\pi/2$ around
$\hat S_{1}$.  Nearly identical results were obtained but on alternative
Stokes parameters.  Fig.~\ref{rotatefig} a) shows Poincar\'{e} sphere
representations of this rotation for the polarization states produced
by two amplitude squeezed beams. In Fig.~\ref{rotatefig} b) the combination of 
a amplitude squeezed vacuum and a bright coherent beam exemplifies that different
orientations of the noise volume can be generated using appropriate combination of 
waveplates.

\section{Channel Capacity of Polarization Squeezed Beams}

The reduced level of fluctuations in polarization squeezed light can
be used to improve the channel capacity of communication protocols.
Let us consider information encoded on the sidebands of a bandwidth
limited single spatial mode laser beam.  We assume that only direct
detection is employed, or in other words, that phase sensitive
techniques such as homodyne measurement are not available.  This is
not an artificial constraint since phase sensitive techniques are
technically difficult to implement and are rarely utilized in
conventional optical communications systems.  

An upper bound to the
amount of information that can be carried by a bandwidth limited
additive white Gaussian noise channel is given by the Shannon capacity C \cite{shan} in bits per dimension
\begin{equation}
C={1 \over 2} \log_{2} (1+R) ~.
\label{sc}
\end{equation}
$R$ is the signal to noise ratio of the channel and is given by the ratio of the spectral variance of the signal modulation $V_s$ and the noise spectral variance $V_n$
\begin{equation}
R = \frac{V_s}{V_n} ~.
\label{R}
\end{equation}
We wish to compare the channel capacities achieveable with pure coherent 
and squeezed states for a given average photon number in the
sidebands $\bar n$ where $\bar n \!=\!  \langle \delta \hat
a_{H}^{\dagger} \delta \hat a^{ }_{H} \!  + \!  \delta \hat
a_{V}^{\dagger} \delta \hat a^{ }_{V} \rangle $.  Note that $\bar n$ takes into account both, the signal modulation and the squeezing. An overview of quantum noise limited channel capacities may be found in \cite{YHa86} and \cite{CDr94}.

First let us consider strategies which might be employed with a coherent light
beam.  In conventional optical communication systems the polarization
degrees of freedom are ignored completely and information is encoded
only on $\hat S_{0}$ as intensity fluctuations. 
Taking $\alpha_{V}\!=\!0$ the variance of the Stokes operator $\hat S_{0}$ is given by $V_{0}\!=\!\alpha_{H}^{2}(V_{s}+1)$ in accordance with Eq.~(\ref{var2}).
For this one-dimensional coherent channel $V_n \!=\! \langle |\delta X_{H}^{+}|^{2}\rangle \!=\! 1$ and therefore $R \!=\! V_s$.
For this arrangement it can be shown that the average photon number per bandwidth per second is
$\bar n \!=\!\frac{1}{4} V_{s}$ providing a photon resource limited Shannon capacity of
\begin{equation}
C_{\rm coh}^{i} = {1 \over 2} \log_{2} (1+4 \bar n) ~,
\label{c1}
\end{equation}
as a function of $\bar n$.  This is a non-optimal strategy however.
Examining Eqs.~(\ref{Scomrel})~and~(\ref{expec}) we see that it is
possible to choose an arrangement for which two of the Stokes
operators commute and so can be measured simultaneously.  Indeed it is
easy to show that such simultaneous measurements can be made using
only linear optics and direct detection.  In particular let us assume
that $\alpha_{H}\!=\!\alpha_{V}$ such that $\hat S_{2}$ and $\hat S_{3}$
commute, and use $\hat S_{2}$ and $\hat S_{3}$ as two independent
information channels. Then $\hat S_{2}\!=\!\hat S_{0}$ and the information in both dimensions can be simultaneously extracted by subtracting and adding the photocurrents of the same pair of detectors.  
Assuming equal signal to noise ratios
$R_{2}\!=\!R_{3}$ we find that $V_{s2}\!=\!V_{s3}\!=\!2\bar n$, and the channel
capacity may be written
\begin{eqnarray}
C_{\rm coh}^{ii} & = & {1 \over 2} \log_{2} (1+R_{2})+{1 \over 2}
\log_{2} (1+R_{3}) \nonumber \\ & = & \log_{2} ({1+2\bar n}) ~.
\label{c2}
\end{eqnarray}
This channel capacity is always greater than that of Eq.~(\ref{c1})
and for large $\bar n$ is 100~\% greater.  

\begin{figure}[h!]
  \vspace{3mm}
  \centerline{\includegraphics[width=8.5cm]{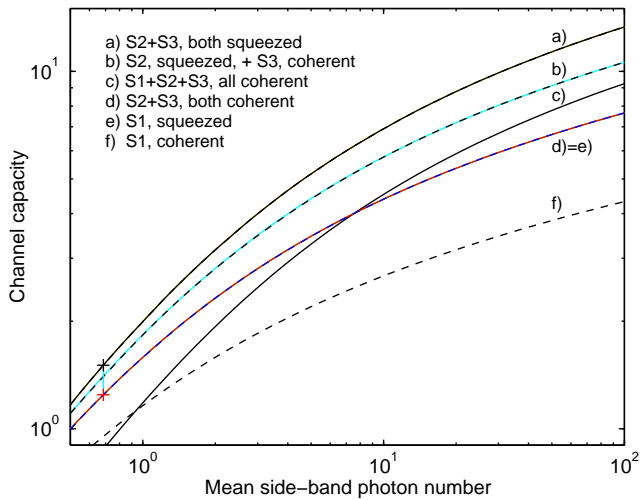}}
  \vspace{0mm}
  \caption{Calculated channel capacities for various continuous variable 
	polarization states. The channels dimension and the channels quantum noise performance is varied. The cross in the upper curve marks the channel capacity which can be achieved using the polarization squeezed state in Fig.~\ref{AmpResults} and is compared with the optimum coherent state channel capacity (lower cross).}
  \label{channelcapacity}
\end{figure}

For sufficiently high $\bar
n$ a further improvement in channel capacity can be achieved.
Consider placing signals on all three Stokes operators.  Because of
the non-commutation of $\hat S_{1}$ with $\hat S_{2}$ and $\hat S_{3}$
it is not possible to read out all three signals without a measurement
penalty.  Suppose the receiver adopts the following strategy: divide
the beam on a beamsplitter with transmitivity $\epsilon$ and then
measure $\hat S_{1}$ on the reflected output and $\hat S_{2}$ and
$\hat S_{3}$ on the other output.  Division of the beam will reduce
the measured signal to noise ratios due to the injection of quantum
noise at the beamsplitter such as $R_{1}\!=\!(1-\epsilon)V_{s1}$,
$R_{2}\!=\!\epsilon V_{s2}$ and $R_{3}\!=\!\epsilon V_{s3}$.  We find that 
for large $\bar n$ an
optimum is reached with $\epsilon\!=\!2/3$ and the signal photon number in
each Stokes parameter being $\bar n_{1}\!=\!\bar n_{2}\!=\!\bar n_{3}\!=\!\bar
n/3$. Hence the channel capacity is
\begin{eqnarray}
C_{\rm coh}^{iii} & = & {1 \over 2} \log_{2} (1\!+\!R_{1})+ {1 \over 2}
\log_{2} (1\!+\!R_{2})+{1 \over 2} \log_{2} (1\!+\!R_{3}) \nonumber \\
& = & {1 \over 2} \log_{2} ({1+{4 \over 9}\bar n}) + \log_{2} ({1+{8
\over 9}\bar n}) ~.
\label{c3}
\end{eqnarray}
This capacity beats that of Eq.~(\ref{c2}) for $\bar n \!  > \! 7.56$. 
In summery the optimum coherent channel capacity is given by Eq.~(\ref{c3}) for average photon numbers $\bar n \! > \! 7.56$ and by Eq.~(\ref{c2}) for lower values, see Fig.~\ref{channelcapacity} traces c) and d). 

Now let us examine the effect of polarization squeezing on the channel
capacity.  Consider first the simple case of intensity modulation on a
single squeezed beam, which is equivalent to using a squeezed beam in
the first coherent case considered (case i).
The channel capacity can be maximized by optimizing the fraction of 
photons which are introduced by squeezing the quantum noise and the residual
fraction of photons which actually carries the signal. For large photon numbers 
we find a proportioning of 0.5. For an average photon number of 1 just 1/3 of 
that photon should be used to reduce the quantum noise.
The maximum channel capacity for a one-dimensional squeezed channel is found 
to be 
\begin{eqnarray}
C_{\rm 1sqz}^{i} & = & \log_{2} ({1+2\bar n}) ~,
\end{eqnarray}
which was previously given in \cite{YHa86}.
This capacity beats only the corresponding coherent state. It is as
efficient as the two-dimensional coherent channel, but less efficient 
than the three-dimensional coherent one for large photon numbers.

Consider now a polarization squeezed beam which is produced from a
minimum uncertainty squeezed beam and a coherent beam, as in Section
\ref{exp1}.  Suppose, as in case ii, that $\hat S_{2}$ and $\hat
S_{3}$ commute and arrange that $\hat S_{2}$ has fluctuations at the
quantum noise level while $\hat S_{3}$ is optimally squeezed.  Again
signals are encoded on $\hat S_{2}$ and $\hat S_{3}$.  The channel
capacity can be maximized by adjusting the relative signal sizes on
the two Stokes operators for fixed average photon number as a function
of the squeezing.  Here 1/3 of the photons is used to squeeze and 2/3
is split equally for the two dimensions.
The resultant maximum channel capacity is
\begin{eqnarray}
C_{\rm 1sqz}^{ii} & = & {3 \over 2} \log_{2} ({1+{4 \over 3}\bar n}) ~.
\label{s1}
\end{eqnarray}
This always beats all three coherent state cases considered here, but
in the limit of large $\bar n$ the advantage is minimal since the
scaling with photon number is the same as that of $C_{\rm
coh}^{iii}$ in Eq.~(\ref{c3}).  

If the polarization squeezed beam is
produced from two amplitude squeezed beams as in Section \ref{exp2}
the enhancement becomes more significant.  Suppose again that $\hat
S_{2}$ and $\hat S_{3}$ commute but now that both are optimally
squeezed.  Again encoding on $\hat S_{2}$ and $\hat S_{3}$, and
varying the signal strength as a function of squeezing to maximize the
channel capacity for a given $\bar n$. The maximum is reached when the 
photons are used to squeeze the noise and transport information in equal shares.
The channel capacity for this arrangement is given by
\begin{eqnarray}
C_{\rm 2sqz}^{ii} & = & 2 \log_{2} ({1+\bar n}) ~,
\label{s2}
\end{eqnarray}
which for large $\bar n$ is 33\% greater than both the optimum
coherent scheme and the scheme using a single quadrature squeezed
beam. No further improvement of the channel capacity can be obtained 
by encoding the information on three Stokes parameters, as in case iii.
Optimization of the beam splitter reflectivity results in the already 
considered two-dimensional arrangement. This is not a surprising result 
since the third Stokes parameter is anti-squeezed. 
Fig.~\ref{channelcapacity} summarizes our results.

Finally we assess the channel capacities that could in
principle be achieved using the polarization squeezed state generated
in our experiment from two amplitude squeezed beams.  
The polarization squeezing achieved in Fig.~\ref{AmpResults} implies that, 
in the frequency range of 8~MHz to 10~MHz, 0.17 side-band photons per bandwidth 
and second were present in each of the two dimensions. This is an optimum 
quantum resource to transmit 0.68 side-band photons. 
Signals sufficiently high above detector dark noise would achieve a channel 
capacity that is around 21\% 
greater than the ideal channel capacity achievable from a coherent beam with 
the same average side-band photon number (see crosses in Fig.~\ref{channelcapacity}).

\section{Conclusion}

The field of quantum communication and computation is receiving much
attention. The continuous variable polarization states investigated here, are
one of the most promising candidates for carrying the information in a
quantum network. In this paper we have characterized the non classical
properties of these states on the basis of the Stokes operators and their
variances. Different classes of polarization squeezed states have been
generated and experimentally characterized. We compared the coherent
polarization state in Fig.$\!$~\ref{BallsAll}a) with squeezed polarization 
states generated from a
single amplitude squeezed beam Fig.$\!$~\ref{BallsAll}c) and from two amplitude squeezed beams 
Fig.$\!$~\ref{BallsAll}e), and proved that
squeezing of better than 3~dB of three Stokes parameters ($\hat S_{0}$,
$\hat S_{1}$, and $\hat S_{3}$) simultaneously is possible only in the
latter case.
We have theoretically analyzed the channel capacity for several 
communication protocols using continuous variable polarization states. 
For a given average photon number $\bar n$, we found the polarization state 
produced from two quadrature squeezed states can provide a 33\%
greater channel capacity than both the optimum coherent scheme and the
scheme using a single quadrature squeezed beam.


We acknowledge the Alexander von Humboldt foundation for
support of R.~Schnabel; the Australian Research Council for
financial support. This work is a part of EU QIPC Project, 
No.~IST-1999-13071 (QUICOV).


\begin{thebibliography}{12}
\bibitem{footnote1} Here we have excluded the phase difference $\theta$
from the mode operators to illustrate the link to our experiment more clearly.
\end{thebibliography}
\end{document}